\documentclass{article}

\usepackage[final]{neurips_2022_ml4ps}
\usepackage{graphicx}

\usepackage[utf8]{inputenc} % allow utf-8 input
\usepackage[T1]{fontenc}    % use 8-bit T1 fonts
\usepackage{hyperref}       % hyperlinks
\usepackage{url}            % simple URL typesetting
\usepackage{booktabs}       % professional-quality tables
\usepackage{amsfonts}       % blackboard math symbols
\usepackage{nicefrac}       % compact symbols for 1/2, etc.
\usepackage{microtype}      % microtypography
\usepackage{xcolor}         % colors

\title{Plausible Adversarial Attacks on Direct Parameter Inference Models in Astrophysics}

\author{%
  Benjamin Horowitz \\
  Department of Astrophysical Sciences\\
  Princeton University\\
  Princeton, NJ  \\
  \texttt{bhorowitz@princeton.edu} \\
  \And
  Peter Melchior\\
  Department of Astrophysical Sciences\\
  Center for Statistics \& Machine Learning\\
  Princeton University\\
  Princeton, NJ  \\
}

\begin{document}

\maketitle

\begin{abstract}
  In this work we explore the possibility of introducing biases in physical parameter inference models from adversarial-type attacks. In particular, we inject small amplitude systematics into an mixture density networks tasked with inferring cosmological parameters from observed data. The systematics are constructed analogously to white-box adversarial attacks. We find that the analysis network can be tricked into spurious detection of new physics in cases where standard cosmological estimators would be insensitive. This calls into question the robustness of such networks and their utility for reliably detecting new physics.
\end{abstract}

\section{Introduction}

Within the physical sciences, there has been an explosion of interest in direct parameter inference models, i.e. models which seek to map directly from observed data to underlying physical parameters of interest \citep{2017estimatingparam,2019xray,2020dm,2022onegal,2022arXiv220906843S}. Many of these papers make strong claims of superiority of these direct methods compared to standard analysis based on field-level maximum likelihood methods or classical summary statistics.
However, the topic of robustness of these models has not been explored rigorously in the literature. Unknown systematics in experiments that are not included in the training data can, and likely will, significantly affect the inferred parameters in ways that are not apparent to users accustomed to standard analysis methods. The results from direct inference techniques may then be misinterpreted as detection of new physics.

In this work we present one particularly strong pathological example inspired by recent work aiming to map from observed cosmological density fields to underlying physical parameters \citep{2017estimatingparam,2022ApJ...928...44V,2022onegal,2022arXiv220906843S}. Standard power-spectra based methods are known to be sub-optimal because of the non-linear evolution of cosmological density due to gravitational evolution, which forces information from two-point statistics into higher order modes. However, these methods are less sensitive to anomalous noise patterns or systematics because they average over the cosmological fields. Neural networks, on the other hand, particularly those utilizing convolutional architectures, can extract non-linear information beyond what is possible with existing classical summary statistics \citep{2021JCAP...09..039L}. We wonder how robustly they can perform this extraction.

As shown in \citep{2016arXiv160702533K}, adversarial attacks on neural networks can happen in the ``physical world", i.e. in natural images processed via a standard camera, without the need to exactly manipulate individual pixels. In the context of the physical sciences this leads to the question whether conceivable physical systematics, not just specifically crafted pathologies, could result in effective adversarial patterns.
 
In this work, we construct a parameter inference network trained on two dimensional projected dark matter fields. We then construct adversarial attacks via the methods discussed in \citep{2016arXiv160702533K}. We show that there are classes of reasonable systematics that could exist below the noise level of existing experiments, and which existing analysis techniques are insensitive to.

\section{Methodology}

\begin{figure}
    \centering
    \includegraphics[width=1.0\textwidth]{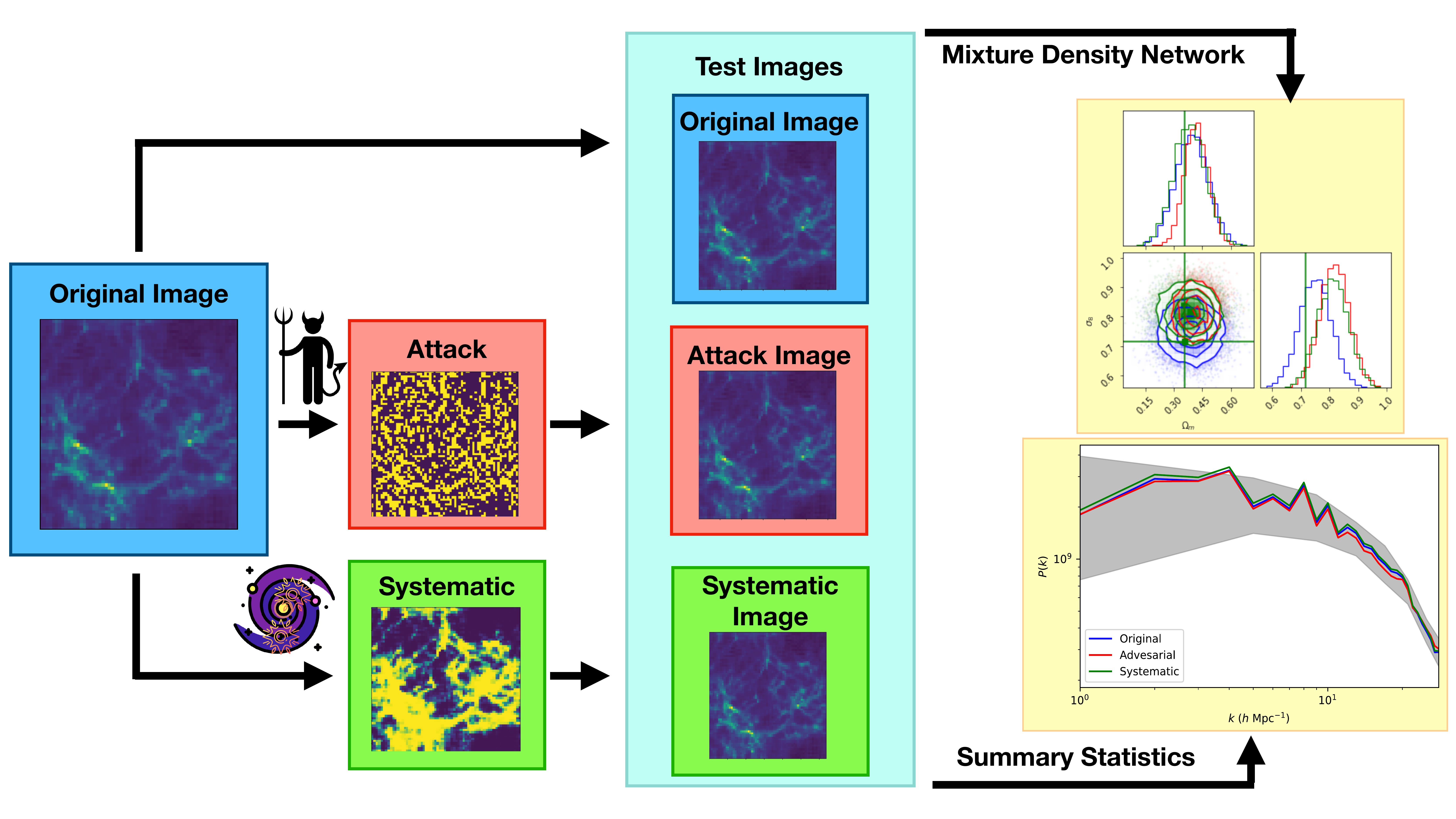}
    \caption{Overview of our workflow to generate and evaluate adversarial-type patterns. From our original image and trained model we can generate an adversarial attack, shown in red, as well as an unknown systematic example, shown in green. Both are added to our original image, leading to nearly imperceptible ($<0.1\%$) changes to a powerspectrum analysis, however both result in significant shifts to the model's inferred cosmological values at a $2\,\sigma$ level. }
    \label{fig:overview_figure}
\end{figure}
\subsection{Simulated training data}
\label{subsec:data}
To simulate idealized astronomical data, we run 10000 small box, particle mesh simulations spanning cosmological parameters $\Omega_M \in [0.05,0.55]$ and $\sigma_8 \in [0.5,1.01]$. For simplicity we use a uniform grid of values with spacing of 0.05 for each parameter value. We use FlowPM \citep{2021A&C....3700505M}, a Tensorflow \citep{2016arXiv160304467A} GPU-based implementation of FastPM \citep{2016MNRAS.463.2273F}. Other that $\Omega_M$ and $\sigma_8$, we hold all other cosmological parameters fixed at the Planck 2015 Best Fit Values \citep{2016A&A...594A..13P}. For our simulations we use a box size of 128 $h^{-1}$Mpc side-length with particle resolution of 64$^3$. These are very coarse simulations by cosmological standards, but are able to capture key differences in the growth of structure caused from variations of the fundamental physics parameters. We use a 70-30 split for training and testing respectively.

\subsection{Network architecture}
\label{subsec:network}
For our model architecture to predict fundamental physics parameters, we will use a convolutional neural network, whose outputs are flattened and passed to a fully-connected mixture density network\citep{bishop1994mixture}. We will aim to predict the possible distributions of our target cosmological parameters ($\Omega_m$ and $\sigma_8$) by predicting the means, $\mu$, standard deviations, $\sigma$, and relative weights, $\pi$ of various Gaussian components, i.e. we assuming a form
\begin{equation}
    p(x|y) = \sum_i^m\pi_i\phi(y|\vec{\theta}_i),
\end{equation}
where $\phi$ are Gaussian distributions defined by parameters $\vec{\theta}_i = (\vec{\mu_i}, \vec{\sigma}_i)$. For this analysis we set $m=1$, i.e. one Gaussian component, although we can easily generalize it to many mixture components. We assume a diagonal covariance for our parameters due to ease of optimization. We make use of the symmetry of our simulated physical system by performing random rotations and translations during training to increase our networks robustness and effective training size. Training was performed on a Tesla V100-PCIE GPU with 32 GB of memory. %For more information about our network architecture see our github repository here: \url{https://github.com/bhorowitz/adversarial-cosmology}.
 
\subsection{Adversarial attack}
\label{subsec:advesarial}

We present two families of adversarial attacks on our network. The first is the ``worst case'', representing the smallest possible permutation resulting in the most significant change in our inferred cosmological parameters. To achieve this we use the Basic Iterative Method (BIM) \citep{2016arXiv160702533K}, an extension of the Fast Gradient Sign Method \citep{2014arXiv1412.6572G}. This is an example of white-box attack where the attacker knows the full likelihood function, $J(\theta,x,y)$, with $\theta$ being the model's (trained) parameters, $x$ the input image and $y$ the simulated true parameters, respectively. The BIM iteration is given by 
\begin{equation}
\label{eq:bim}
    x_{t+1} = \textrm{Clip}(x_{t} + \alpha \times \textrm{Sign}(\nabla_x J(\theta,x_{t},y))),
\end{equation}
where $\textrm{Clip}$ keeps all values below a value of 1. We run this process for ten steps with $\alpha$ value of 0.01. This method will construct the smallest change to the original image that results in the maximum effect on the inferred parameters.

\subsection{Unknown Systematic attack}
\label{subsec:systematic}
For a more realistic example we examine a perturbation which scales as a non-linear function of observed density field. This could, for example, reflect an incorrect calibration of a detector or some unforeseen small-scale hydrodynamical effect in galaxy clusters. We parameterize this model as 
\begin{equation}
\label{eq:sys}
    x_{sys} = x + \beta \times \frac{1}{1+e^{ax+b}}
\end{equation}
To find an ``adversarial" example with this property we use a similar method as BIM but instead of studying the gradient with respect to the underlying field we find it with respect to the parameters of Equation \ref{eq:sys}. Tuning the parameters by hand, we find $a=0.25$ and $b=-15.0$ provides a significant shift to the trained network's accuracy with minimal visual change. We choose $\beta$ such that the total mean squared change induced by our unknown systematic is similar in amplitude to our adversarial attack, finding $\beta=3.0$.

\section{Results}
We train a model as described in Section \ref{subsec:network} using the simulated training data described in \ref{subsec:data}. We stop training after 1000 epochs (approximately one hour), finding suitable accuracy on our test data-set is achieved with minimal marginal improvement in overall loss per additional epoch. We then use the methods described in Sections \ref{subsec:advesarial} and \ref{subsec:systematic} to generate adversarial-type patterns which we add to our test images. We scale our systematic pattern to match the overall change in pixel value of the white-box adversarial pattern. We show this workflow in Figure \ref{fig:overview_figure}.

\begin{figure}
    \centering
    \includegraphics[width=1.0\textwidth]{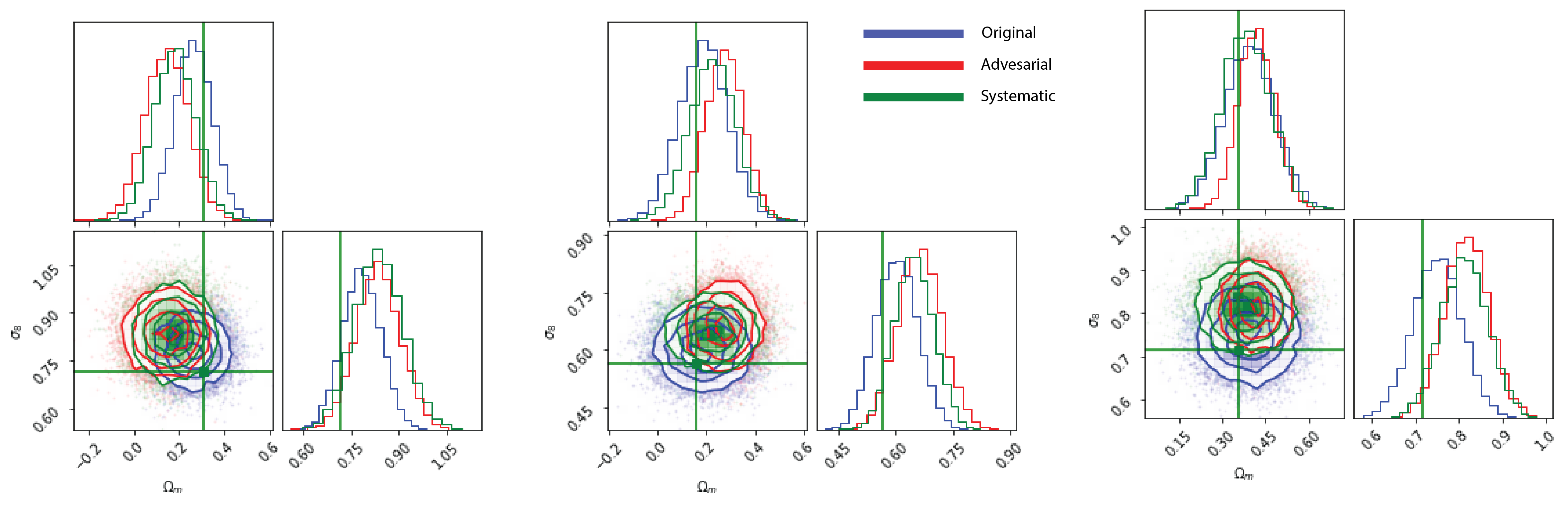}
    \caption{Shown are three random examples from our test data-set, passed through our mixture density network, showing the similarities between our adversarial attack and the mock systematic. While on average the mock systematic results in less of an offset, it still results in a significant bias that could cause a spurious detection. The green circle indicates the simulated truth.}
    \label{fig:examples}
\end{figure}

For comparison, we also calculate the power spectra of the original, adversarial, and systematically altered fields. This is the most common way to extract cosmological information from observed density field data. We find that the adversarial-type patterns generate changes to the power spectra well below the intrinsic cosmological variability (grey bands in Figure \ref{fig:overview_figure}), i.e. this method is robust to small scale unknown systematics. Meanwhile our model is, by construction, highly sensitive to this attack and results in significant nonphysical parameter shifts. We show additional examples of this procedure in Figure \ref{fig:examples}.

Beyond qualitative comparisons, we can compare the resulting distributions based on the average Kullback-Leibler (KL) Divergence. For our test sample, we calculate the KL divergence from the perturbed distribution to the unperturbed distribution (note that the KL Divergence is not symmetric). For our test sample, we find a mean KL divergence of 1.05  (0.98) for the distances adversarial (systematic) attack and the unperturbed example. Meanwhile, the mean symmetrized KL divergence between the adversarial and systematic attacks is 0.67. This indicates our resulting perturbed distributions are on significantly offset from the original distribution on average, while also being similar to each other.

\section{Discussion}

In this work we have constructed a direct parameter inference model for cosmological density fields, which is closely related to those discussed in the literature \citep{2017estimatingparam,2022ApJ...928...44V,2022onegal,2022arXiv220906843S}.
We have shown that these types of neural network estimators are susceptible to adversarial attacks in ways traditional statistics are not. In addition, we have shown that there exists a space of reasonable physical systematics, which, while imperceptible to existing methods, lead to similar biases in cosmological parameters as white-box adversarial attacks. These systematics are closer in distribution, as measured by the KL divergence, to adversarial attacks than to the unperturbed distribution.

We conclude from these tests that there is reason for skepticism about cosmological results drawn from direct parameter inference models. A key feature of these models is that they derive cosmological information from small-scale features in the density fields. Great care is needed to ensure the robustness of such models to all conceivable new systematics, not just to those the community has investigated in the context of traditional analysis methods. Even if doing so leads to reduced constraining power \citep{2017arXiv170403976M,2020arXiv200210716R}. Going forward, we believe it should be standard practice to test parameter inference models with adversarial examples, or, even better, to inject adversarial examples during training to increase the robustness of the final models.

\paragraph{Limitations of work:} In this study, we examined attacks on only one specific network architecture in one context. It is possible other networks would be more or less conducive to this method of attack. For example, neural flow-based models \citep{2022MNRAS.516.2363D} have significantly fewer parameters which could make them more robust to adversarial-type systematics. We also did not explicitly explore examples of adversarial-type attacks to which the power spectrum would be sensitive but neural network models would be not. However, the sensitivity of power spectrum methods is well studied in existing astrophysics literature.

\paragraph{Impact Statement:} This work shows potential limitations of a large (and growing) body of work found throughout the physical sciences which attempts to directly map from observed data to underlying physical parameters. We believe our work should encourages authors of such works to explore the robustness of their models, including potentially incorporating adversarial training. It also elucidates this issue for the broader scientific community to encourage well-founded skepticism on any claimed physical discoveries based on this class of models. 

\bibliographystyle{plainnat}
\bibliography{main}

\begin{thebibliography}{17}
\providecommand{\natexlab}[1]{#1}
\providecommand{\url}[1]{\texttt{#1}}
\expandafter\ifx\csname urlstyle\endcsname\relax
  \providecommand{\doi}[1]{doi: #1}\else
  \providecommand{\doi}{doi: \begingroup \urlstyle{rm}\Url}\fi

\bibitem[{Abadi} et~al.(2016){Abadi}, {Agarwal}, {Barham}, {Brevdo}, {Chen},
  {Citro}, {Corrado}, {Davis}, {Dean}, {Devin}, {Ghemawat}, {Goodfellow},
  {Harp}, {Irving}, {Isard}, {Jia}, {Jozefowicz}, {Kaiser}, {Kudlur},
  {Levenberg}, {Mane}, {Monga}, {Moore}, {Murray}, {Olah}, {Schuster},
  {Shlens}, {Steiner}, {Sutskever}, {Talwar}, {Tucker}, {Vanhoucke},
  {Vasudevan}, {Viegas}, {Vinyals}, {Warden}, {Wattenberg}, {Wicke}, {Yu}, and
  {Zheng}]{2016arXiv160304467A}
Mart{\'\i}n {Abadi}, Ashish {Agarwal}, Paul {Barham}, Eugene {Brevdo}, Zhifeng
  {Chen}, Craig {Citro}, Greg~S. {Corrado}, Andy {Davis}, Jeffrey {Dean},
  Matthieu {Devin}, Sanjay {Ghemawat}, Ian {Goodfellow}, Andrew {Harp},
  Geoffrey {Irving}, Michael {Isard}, Yangqing {Jia}, Rafal {Jozefowicz},
  Lukasz {Kaiser}, Manjunath {Kudlur}, Josh {Levenberg}, Dan {Mane}, Rajat
  {Monga}, Sherry {Moore}, Derek {Murray}, Chris {Olah}, Mike {Schuster},
  Jonathon {Shlens}, Benoit {Steiner}, Ilya {Sutskever}, Kunal {Talwar}, Paul
  {Tucker}, Vincent {Vanhoucke}, Vijay {Vasudevan}, Fernanda {Viegas}, Oriol
  {Vinyals}, Pete {Warden}, Martin {Wattenberg}, Martin {Wicke}, Yuan {Yu}, and
  Xiaoqiang {Zheng}.
\newblock {TensorFlow: Large-Scale Machine Learning on Heterogeneous
  Distributed Systems}.
\newblock \emph{arXiv e-prints}, art. arXiv:1603.04467, March 2016.

\bibitem[Bishop(1994)]{bishop1994mixture}
Christopher~M Bishop.
\newblock Mixture density networks.
\newblock 1994.

\bibitem[{Dai} and {Seljak}(2022)]{2022MNRAS.516.2363D}
Biwei {Dai} and Uro{\v{s}} {Seljak}.
\newblock {Translation and rotation equivariant normalizing flow (TRENF) for
  optimal cosmological analysis}.
\newblock \emph{MNRAS}, 516\penalty0 (2):\penalty0 2363--2373, October 2022.
\newblock \doi{10.1093/mnras/stac2010}.

\bibitem[{Feng} et~al.(2016){Feng}, {Chu}, {Seljak}, and
  {McDonald}]{2016MNRAS.463.2273F}
Yu~{Feng}, Man-Yat {Chu}, Uro{\v{s}} {Seljak}, and Patrick {McDonald}.
\newblock {FASTPM: a new scheme for fast simulations of dark matter and
  haloes}.
\newblock \emph{MNRAS}, 463\penalty0 (3):\penalty0 2273--2286, December 2016.
\newblock \doi{10.1093/mnras/stw2123}.

\bibitem[{Goodfellow} et~al.(2014){Goodfellow}, {Shlens}, and
  {Szegedy}]{2014arXiv1412.6572G}
Ian~J. {Goodfellow}, Jonathon {Shlens}, and Christian {Szegedy}.
\newblock {Explaining and Harnessing Adversarial Examples}.
\newblock \emph{arXiv e-prints}, art. arXiv:1412.6572, December 2014.

\bibitem[{Khosa} et~al.(2020){Khosa}, {Mars}, {Richards}, and {Sanz}]{2020dm}
Charanjit~K. {Khosa}, Lucy {Mars}, Joel {Richards}, and Veronica {Sanz}.
\newblock {Convolutional neural networks for direct detection of dark matter}.
\newblock \emph{Journal of Physics G Nuclear Physics}, 47\penalty0
  (9):\penalty0 095201, September 2020.
\newblock \doi{10.1088/1361-6471/ab8e94}.

\bibitem[{Kurakin} et~al.(2016){Kurakin}, {Goodfellow}, and
  {Bengio}]{2016arXiv160702533K}
Alexey {Kurakin}, Ian {Goodfellow}, and Samy {Bengio}.
\newblock {Adversarial examples in the physical world}.
\newblock \emph{arXiv e-prints}, art. arXiv:1607.02533, July 2016.

\bibitem[{Lazanu}(2021)]{2021JCAP...09..039L}
Andrei {Lazanu}.
\newblock {Extracting cosmological parameters from N-body simulations using
  machine learning techniques}.
\newblock \emph{JCAP}, 2021\penalty0 (9):\penalty0 039, September 2021.
\newblock \doi{10.1088/1475-7516/2021/09/039}.

\bibitem[{Miyato} et~al.(2017){Miyato}, {Maeda}, {Koyama}, and
  {Ishii}]{2017arXiv170403976M}
Takeru {Miyato}, Shin-ichi {Maeda}, Masanori {Koyama}, and Shin {Ishii}.
\newblock {Virtual Adversarial Training: A Regularization Method for Supervised
  and Semi-Supervised Learning}.
\newblock \emph{arXiv e-prints}, art. arXiv:1704.03976, April 2017.

\bibitem[{Modi} et~al.(2021){Modi}, {Lanusse}, and
  {Seljak}]{2021A&C....3700505M}
C.~{Modi}, F.~{Lanusse}, and U.~{Seljak}.
\newblock {FlowPM: Distributed TensorFlow implementation of the FastPM
  cosmological N-body solver}.
\newblock \emph{Astronomy and Computing}, 37:\penalty0 100505, October 2021.
\newblock \doi{10.1016/j.ascom.2021.100505}.

\bibitem[{Ntampaka} et~al.(2019){Ntampaka}, {ZuHone}, {Eisenstein}, {Nagai},
  {Vikhlinin}, {Hernquist}, {Marinacci}, {Nelson}, {Pakmor}, {Pillepich},
  {Torrey}, and {Vogelsberger}]{2019xray}
M.~{Ntampaka}, J.~{ZuHone}, D.~{Eisenstein}, D.~{Nagai}, A.~{Vikhlinin},
  L.~{Hernquist}, F.~{Marinacci}, D.~{Nelson}, R.~{Pakmor}, A.~{Pillepich},
  P.~{Torrey}, and M.~{Vogelsberger}.
\newblock {A Deep Learning Approach to Galaxy Cluster X-Ray Masses}.
\newblock \emph{APJ}, 876\penalty0 (1):\penalty0 82, May 2019.
\newblock \doi{10.3847/1538-4357/ab14eb}.

\bibitem[{Planck Collaboration}(2016)]{2016A&A...594A..13P}
{Planck Collaboration}.
\newblock {Planck 2015 results. XIII. Cosmological parameters}.
\newblock \emph{AAP}, 594:\penalty0 A13, September 2016.
\newblock \doi{10.1051/0004-6361/201525830}.

\bibitem[{Raghunathan} et~al.(2020){Raghunathan}, {Xie}, {Yang}, {Duchi}, and
  {Liang}]{2020arXiv200210716R}
Aditi {Raghunathan}, Sang~Michael {Xie}, Fanny {Yang}, John {Duchi}, and Percy
  {Liang}.
\newblock {Understanding and Mitigating the Tradeoff Between Robustness and
  Accuracy}.
\newblock \emph{arXiv e-prints}, art. arXiv:2002.10716, February 2020.

\bibitem[{Ravanbakhsh} et~al.(2017){Ravanbakhsh}, {Oliva}, {Fromenteau},
  {Price}, {Ho}, {Schneider}, and {Poczos}]{2017estimatingparam}
Siamak {Ravanbakhsh}, Junier {Oliva}, Sebastien {Fromenteau}, Layne~C. {Price},
  Shirley {Ho}, Jeff {Schneider}, and Barnabas {Poczos}.
\newblock {Estimating Cosmological Parameters from the Dark Matter
  Distribution}.
\newblock \emph{arXiv e-prints}, art. arXiv:1711.02033, November 2017.

\bibitem[{Shao} et~al.(2022){Shao}, {Villaescusa-Navarro},
  {Villanueva-Domingo}, {Teyssier}, {Garrison}, {Gatti}, {Inman}, {Ni},
  {Steinwandel}, {Kulkarni}, {Visbal}, {Bryan}, {Angles-Alcazar}, {Castro},
  {Hernandez-Martinez}, and {Dolag}]{2022arXiv220906843S}
Helen {Shao}, Francisco {Villaescusa-Navarro}, Pablo {Villanueva-Domingo},
  Romain {Teyssier}, Lehman~H. {Garrison}, Marco {Gatti}, Derek {Inman},
  Yueying {Ni}, Ulrich~P. {Steinwandel}, Mihir {Kulkarni}, Eli {Visbal},
  Greg~L. {Bryan}, Daniel {Angles-Alcazar}, Tiago {Castro}, Elena
  {Hernandez-Martinez}, and Klaus {Dolag}.
\newblock {Robust field-level inference with dark matter halos}.
\newblock \emph{arXiv e-prints}, art. arXiv:2209.06843, September 2022.

\bibitem[{Villaescusa-Navarro} et~al.(2022{\natexlab{a}}){Villaescusa-Navarro},
  {Ding}, {Genel}, {Tonnesen}, {La Torre}, {Spergel}, {Teyssier}, {Li},
  {Heneka}, {Lemos}, {Angl{\'e}s-Alc{\'a}zar}, {Nagai}, and
  {Vogelsberger}]{2022onegal}
Francisco {Villaescusa-Navarro}, Jupiter {Ding}, Shy {Genel}, Stephanie
  {Tonnesen}, Valentina {La Torre}, David~N. {Spergel}, Romain {Teyssier}, Yin
  {Li}, Caroline {Heneka}, Pablo {Lemos}, Daniel {Angl{\'e}s-Alc{\'a}zar},
  Daisuke {Nagai}, and Mark {Vogelsberger}.
\newblock {Cosmology with One Galaxy?}
\newblock \emph{APJ}, 929\penalty0 (2):\penalty0 132, April 2022{\natexlab{a}}.
\newblock \doi{10.3847/1538-4357/ac5d3f}.

\bibitem[{Villaescusa-Navarro} et~al.(2022{\natexlab{b}}){Villaescusa-Navarro},
  {Wandelt}, {Angl{\'e}s-Alc{\'a}zar}, {Genel}, {Manuel Zorrilla Matilla},
  {Ho}, and {Spergel}]{2022ApJ...928...44V}
Francisco {Villaescusa-Navarro}, Benjamin~D. {Wandelt}, Daniel
  {Angl{\'e}s-Alc{\'a}zar}, Shy {Genel}, Jose {Manuel Zorrilla Matilla},
  Shirley {Ho}, and David~N. {Spergel}.
\newblock {Neural Networks as Optimal Estimators to Marginalize Over Baryonic
  Effects}.
\newblock \emph{APJ}, 928\penalty0 (1):\penalty0 44, March 2022{\natexlab{b}}.
\newblock \doi{10.3847/1538-4357/ac54a5}.

\end{thebibliography}

\end{document}